\documentclass[aip,pop,psfig,twocolumn,showpacs,superscriptaddress,reprint]{revtex4-1}
\usepackage{color}
\usepackage{graphics}
\usepackage{epsfig}
\usepackage{ulem}
\usepackage{cancel}

\begin{document}
\title{Wakes and precursor soliton excitations by a moving charged object in a plasma}
\author{Sanat Kumar Tiwari}
\email{sanat-tiwari@uiowa.edu}
\affiliation{Institute for Plasma Research, Bhat , Gandhinagar - 382428, India }
\affiliation{Department of Physics and Astronomy, University of Iowa, Iowa City, Iowa 52242, USA}
\author{Abhijit Sen}
\email{senabhijit@gmail.com}
\affiliation{Institute for Plasma Research, Bhat , Gandhinagar - 382428, India }
\date{\today}
\begin{abstract}
We study the evolution of nonlinear ion acoustic wave excitations due to a moving charged source in a plasma. Our numerical investigations of the full set of cold fluid equations goes beyond the usual weak nonlinearity approximation and shows the existence of a rich variety of solutions including wakes, precursor solitons and ``pinned'' solitons that travel with the source velocity. These solutions represent a large amplitude generalization of solutions obtained in the past for the forced Korteweg deVries equation and can find useful applications in a variety of situations in the laboratory and in space wherever there is a large relative velocity between the plasma and a charged object.\\
\end{abstract}
\maketitle
\section{Introduction}
\label{intro}
The Korteweg - deVries (KdV) equation \cite{Korteweg1895} has for long served as a paradigmatic model for the study of low frequency weakly nonlinear dispersive waves in a wide variety of physical systems \cite{Miura1976}. As a completely integrable nonlinear partial differential equation the model has not only been the subject of numerous mathematical studies \cite{Zabusky1965,Washimi1966,Gardner1967,Miura1968,Miles1981} but has also lent itself to practical applications due to its large class of exact nonlinear solutions \cite{Zabusky1971}. In particular, the class of solutions consisting of nonlinear solitary pulses, known as solitons, have been much studied both theoretically and experimentally because of their remarkable properties. In plasma physics the KdV model has been very successfully applied to the study of nonlinear ion-acoustic waves \cite{Washimi1966,Ikezi1973,Nishikawa1974,Watanabe1978,Nagasawa1981,Nakamura1996,Tran2007,Lonngren1998} as well as to other low frequency waves such as dust ion-acoustic waves (DAWs) in dusty plasma systems \cite{Mamun1996,Ma1997,shukla_njp_03,merlino_12,sanat_weak_njp,pintu_prl}. The soliton form of ion acoustic waves results from an exact balance of the nonlinear steepening and dispersive broadening effects that a finite wave pulse experiences in the medium. A similar mechanism is responsible for the formation of solitonic structures in surface water waves - the classic historical example of which is the long lived mound of water that Sir Russell Scott had observed in a canal and chased for long on horseback \cite{russell_sol}. The soliton structure in this case had resulted from the sudden stopping of a boat that was being dragged in the canal and that had caused a large amplitude perturbation of the water surface. More recently water wave solitons have also been observed and studied for a situation where the boat keeps moving at a certain speed and acts as a continuous source of perturbation \cite{Akylas1984,Ertekin1986,Lee1989}. These driven solitons are found to be generated ahead of the boat when the boat velocity exceeds the phase velocity of the linear surface waves while wake waves are seen to be generated behind the boat for all velocities. This fascinating phenomenon of precursor solitons has received much attention in the hydrodynamic literature and has been well explained by a driven version of the KdV model, namely the forced KdV (fKdV) equation \cite{Wu1987,Smythe1990,Binder2014}.  Surprisingly, to the best of our knowledge, the fKdV equation has not received any attention in the plasma community although physical situations analogous to that in hydrodynamics are conceivable in plasma physics. The direct equivalent of the moving object in a neutral fluid is a moving charged object in a plasma that can give rise to ion acoustic wave excitations in the form of wakes and or solitons. While wake fields due to moving charged objects are well known in plasmas \cite{Neufeld1955,Rostoker1961,Sylla2011,Hutchinson2011,Malka2012,Block2012,Ali2013} the phenomenon of excitation of precursor solitons ahead of the moving charge has not been investigated. In a recent theoretical study we have derived the fKdV equation as an appropriate model equation for the nonlinear driven excitation of ion acoustic waves in a plasma and discussed the possibility of creating precursor solitons in the ionospheric plasma \cite{Sen2015}. Charged objects moving at supersonic speeds can also occur in laboratory situations such as in ion beam driven inertial fusion schemes. The excitation of precursor solitons in such a case can impact the nature of energy deposition on the target pellet and thereby influence its compression and heating characteristics. It is not always necessary for the object to move at supersonic speeds but a fluid (plasma) moving at a supersonic speed over a stationary (charged) object can also give rise to a similar scenario. Examples of such fast streaming plasmas are plasma jets emitted from astrophysical objects and, nearer home, the solar wind hitting the earth. Supersonic plasma flows can also be easily created in table top laboratory experiments and can offer a means to test out the existence of such a phenomenon in plasmas. Thus it is important to understand the conditions under which such excitations can occur and the physical mechanisms underlying such a phenomenon.

In the present paper we show that the excitation of precursor solitons is not restricted to the fKdV equation but can also occur within the general framework of the cold ion fluid equations that is so extensively used to study ion acoustic waves in plasmas. Our numerical investigation of the full set of fluid equations, that are driven by a source term, reveal a rich variety of nonlinear solutions some of which are generalizations of the fKdV equation and include wakes, precursor solitons and pinned solitons. We delineate the existence regions of these solutions and also investigate their characteristics as a function of the driver velocity, the driver amplitude and the width of the driving source. The paper is organized as follows. In the next section, section \ref{gov}, we describe the one dimensional cold fluid model with a moving source term and briefly discuss its reduction in various limits to the KdV, the generalized KdV and the fKdV equations. In section \ref{ext} we look for stationary frame solutions of the system that are moving at the speed of the source term and discuss their relation with the soliton solutions of the generalized Sagdeev potential (in the absence of the source) and the so called ``pinned'' solitons of the fKdV in the presence of the source. In section \ref{rand_exact} we present our numerical solutions of the full set of fluid equations that identify conditions under which we get precursor solitons and wakes. Section \ref{concl} provides a summary of our results and a brief discussion of their potential significance for plasma applications.

\begin{figure}[ht]
\centering
\includegraphics[height=7.0cm,width=8.0cm]{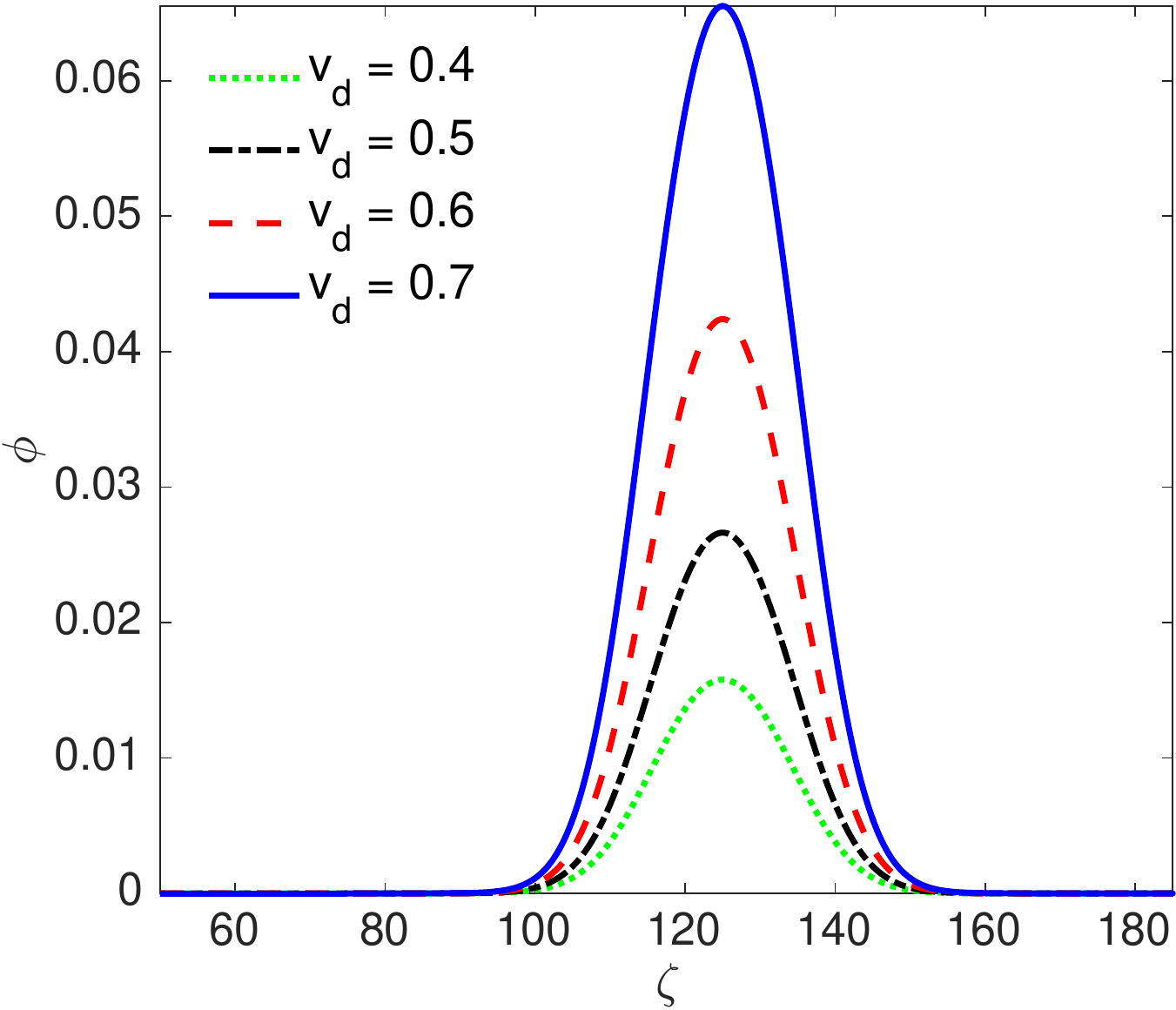}
\caption{Exact localized (``pinned'') solutions $\phi$ of Eq.~\ref{fn_diff2} with the source term $S(\zeta) = A_m \exp\left[-(\zeta/\delta)^2\right]$ with 
 $\delta = 12.0$, $A_m = 0.1$ for different values of the moving frame velocity, namely, 
 $v_d = 0.4,0.5,0.6$ and $0.7$ respectively. These solutions move at the frame velocity and are thus
 stationary solutions in the frame of the moving source.}
\label{figure_1}
\end{figure}
\section{Model equations}
\label{gov}
The evolution of finite amplitude ion acoustic waves in a plasma are well described by a basic fluid model that takes account of the ion dynamics and treats the lighter electron species to have a Boltzmann distribution. We consider a situation where the ion acoustic wave excitations are being continuously driven by a charged source moving at a constant velocity $v_d$. For simplicity we consider a one dimensional situation for which the model equations can be written down as,
\begin{eqnarray}
&& \frac{\partial n}{\partial t} + \frac{\partial (n u)}{\partial x} = 0 \label{basiceq1} \\
&& \frac{\partial u}{\partial t} + u \frac{\partial u}{\partial x} + \frac{\partial \phi}{\partial x} = 0  \label{basiceq2} \\
&& \frac{\partial^2 \phi}{\partial x^2} = e^{\phi} - n + S(x - v_d t)
 \label{basiceq3}
\end{eqnarray}
Here, $n,u$ are the density and velocity of the ion fluid respectively and $\phi$ is the electrostatic potential. Equation~(\ref{basiceq1}) is the ion continuity equation, Eq.
~(\ref{basiceq2}) is the ion momentum equation and Eq.~(\ref{basiceq3}) is Poisson's equation where $S$ is the additional charge density arising from the moving charged source. The above dimensionless equations have the following normalization:
\begin{eqnarray}
&& x \rightarrow x/\lambda_{De}; \hspace{0.4cm} t \rightarrow \frac{c_s}{\lambda_{De}}t; \hspace{0.4cm} n \rightarrow \frac{n}{n_0}; \nonumber \\
&& \hspace{0.2cm} u \rightarrow \frac{u}{c_s}; \hspace{0.6cm} \phi \rightarrow \frac{e\phi}{k_BT_{e}}
 \nonumber
\end{eqnarray}
where $\lambda_{De}$ is the electron Debye length, $c_s$ is the ion acoustic velocity, $T_{e}$ is the electron temperature, $k_B$ is the Boltzmann constant, $e$ is the electronic charge and the ions are treated as being cold ($T_{i}=0$). In the absence of the source term, a reductive perturbation analysis of Eqs.~(\ref{basiceq1}-\ref{basiceq3}) leads to the well known Korteweg-deVries equation that has been the subject of numerous studies in the past. The KdV equation, a completely integrable nonlinear partial differential equation admits a special class of exact solutions called solitons that arise from a perfect balance between nonlinear steepening and dispersive broadening of a finite amplitude pulse. Ion acoustic solitons have received a great deal of theoretical as well as experimental attention in plasma physics and have been invoked to explain diverse nonlinear phenomena observed in the laboratory as well as in space plasmas. The KdV equation is appropriate for small amplitude waves such that the approximation of weak nonlinearity invoked for the perturbation analysis holds good \cite{Washimi1966}. Such an assumption permits the expansion of the exponential term in Poisson's equation and the neglect of terms beyond $\phi^2$ to close the system. Using the same approximation and assuming the source term $S$ in Eq.~(\ref{basiceq3}) to be of order $\varepsilon^2$ (where $\varepsilon$ is an expansion parameter that represents the smallness of the amplitude) the reductive perturbation technique can be employed to derive a driven version of the 
KdV equation, the so-called forced KdV (fKdV) equation, given by,

\begin{equation}
\frac{\partial \phi_1}{\partial \tau} +\phi_1 \frac{\partial \phi_1}{\partial \xi} +\frac{1}{2}\frac{\partial^3 \phi_1}{\partial \xi^3} = \frac{1}{2}\frac{\partial S(\xi +Ft)}{\partial \xi}
\label{fKdv}
\end{equation}
where
$$ \xi = \varepsilon^{1/2} (x - u_{ph} t) \;\;;\;\; \tau = \varepsilon^{3/2} t $$
are stretched variables and $S(\xi +Ft)$ is the source term, with $F=1 -v_d$ (note the unity term stands for the normalized linear phase velocity of the ion acoustic). For a detailed derivation of Eq.~(\ref{fKdv}) the reader is referred to reference \cite{Sen2015}. Although not as well known or widely studied as the KdV equation the fKdV equation has nevertheless been the subject of numerous studies for its interesting mathematical properties as well as for its applications in hydrodynamics \cite{Akylas1984,Ertekin1986,Lee1989}. It admits a class of exact solitonic solutions for some special analytic forms of the source term. These solutions move at the same speed as the source and are called ``pinned'' solitons. For arbitrary forms of the source function and when
the source moves at supersonic velocities the solution of the fKdV shows the continuous periodic emission of solitons ahead of the source and the generation of weak wake fields behind the source. These precursor solitons move faster than the source and are a novel feature of the fKdV equation. They have been well studied in the fluid mechanics community and have been invoked to explain experimental observations of nonlinear water wave structures created ahead of a ship moving in a narrow channel or in situations when a fluid flows over a submerged object \cite{Wu1987}.

\begin{figure}[ht]
\centering
\includegraphics[height=7.0cm,width=9.0cm]{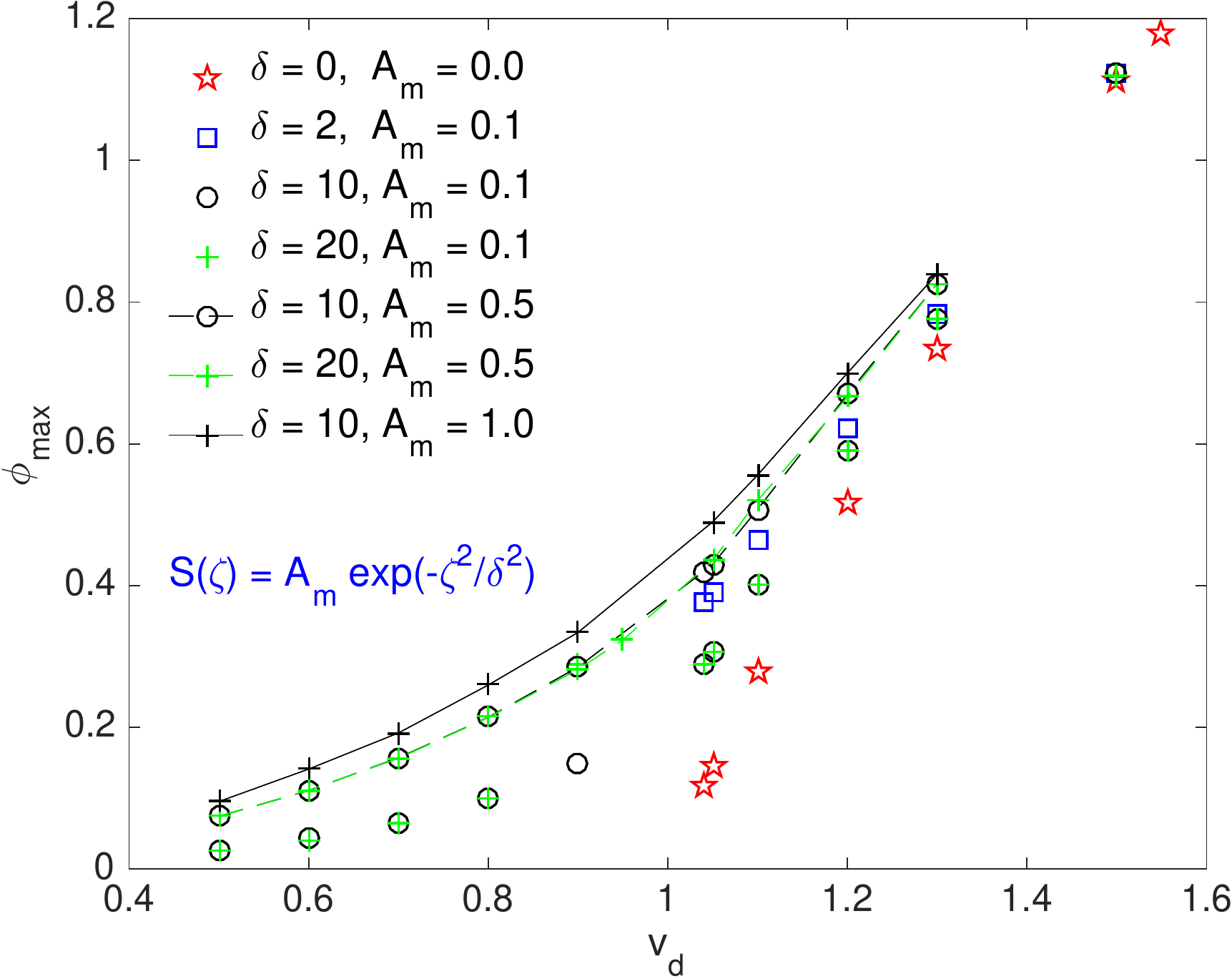}
\caption{A plot in the $\phi_{max}$ vs $v_d$ phase space showing the existence regions of ``pinned'' solitons. Here $\phi_{max}$ is the maximum amplitude of the soliton and $v_d$ is the common velocity of the source and soliton. For comparison soliton solutions of the KdV equation (i.e. in the absence of the source term in Eq.~\ref{fn_diff2}), represented by star markers are also shown where $v_d$ in this case is to be interpreted as the velocity of the soliton. A marked difference to be noted for the two cases is that whereas the KdV solitons can exist
only for $v_d>1$ the ``pinned'' solitons (driven by a source) can arise for both $v_d<1$ as well as for $v_d>1$.}
\label{phi_max_vs_vd}
\end{figure}
\begin{figure*}[ht]
\centering
\includegraphics[height=5.0cm,width=16.0cm]{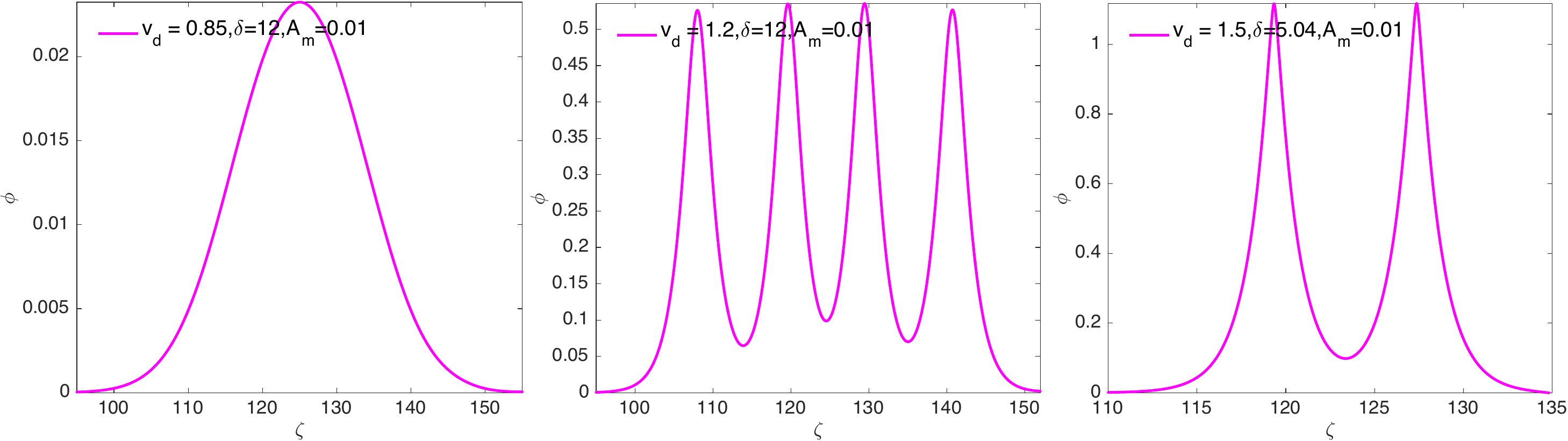}
\caption{A variety of ``pinned'' solitons for $v_d>1$.}
\label{fig_vd_gt1}
\end{figure*}

\section{Stationary frame solutions}
\label{ext}
Before studying the arbitrary time dependent driven solutions of Eqs.~(\ref{basiceq1}-\ref{basiceq3}) let us first examine the possibility of stationary solutions of the system that move with the velocity of the source. In other words we would like to look at generalizations of the pinned soliton solutions of the fKdV for arbitrary amplitude excitations. We therefore transform Eqs.~(\ref{basiceq1}-\ref{basiceq3}) to a moving frame by employing the set of coordinates $\zeta = x - v_d t$ and $t = \tau$.
For stationarity we then require $\partial/\partial \tau = 0$ and $\partial/\partial x = \partial/\partial \zeta$.
Equations~(\ref{basiceq1}-\ref{basiceq3}) then take the form:
\begin{eqnarray}
&& v_d-u = \frac{v_d}{n}   \label{eqst1} \\
&& \phi = u v_d - \frac{u^2}{2}  \label{eqst2} \\
&& \frac{\partial^2 \phi}{\partial \zeta^2} = e^{\phi} - n + S(\zeta)
 \label{eqst3}
\end{eqnarray}
Using (\ref{eqst1}) and (\ref{eqst2}) we can write Eq.~(\ref{eqst3}) entirely in terms of the variable $\phi$ as
\begin{equation}
 \frac{\partial^2 \phi}{\partial \zeta^2} = e^{\phi} -\frac{v_d}{\sqrt{v_{d}^2 - 2 \phi}}  + S(\zeta)
 \label{fn_diff2}
\end{equation}
Note that in the absence of the source term ($S(\zeta)=0$), Eq.~(\ref{fn_diff2}) can be integrated once to get
\begin{equation}
\frac{1}{2} \left( \frac{\partial \phi}{\partial \zeta} \right)^2  - e^{\phi} - v_d \sqrt{v_d^2 - 2\phi} + C =0
\label{sag_eq}
\end{equation}

\noindent
where the constant $C=(1 + v_d^2)$ for solitonic solutions that have boundary conditions $\phi \rightarrow 0$
and $d\phi/d\zeta \rightarrow 0$ as $\zeta \rightarrow \pm \infty$. Eq.~(\ref{sag_eq}) is in the classical form where the first term can be identified as the ``kinetic energy'' of a pseudo-particle moving in the ``pseudo-potential'' $V(\phi,v_d)$ given by,
\begin{equation}
\label{sag_p}
V(\phi,v_d) = 1 +v_d^2 - e^{\phi}- v_d \sqrt{v_d^2 - 2\phi}
\end{equation} 
An analysis of this so-called Sagdeev potential, a much explored tool, enables us to go beyond the weak non-linearity limit and determine the existence regions for larger amplitude solitons as a function of the soliton Mach number. It also helps us to determine the maximum possible amplitude of such solitons, $\phi_M$, by demanding that 
\begin{equation}
e^{\phi_M} + v_d \sqrt{v_d^2 - 2\phi_M} -(1 + v_d^2) = 0
\label{sag_eq_M}
\end{equation}
The maximum amplitude is achieved as $v_d^2 \rightarrow 2\phi_M$ which converts Eq.~(\ref{sag_eq_M}) into a transcendental equation in $\phi_M$ and whose numerical solution yields $\phi_M =1.2564$. Thus the upper limit
on $v_d$ is $\sqrt{2\phi_M} = 1.58518$. Soliton solutions do not exist beyond this value of the Mach number and also for values of $v_d$ less than unity. In the presence of the source term these existence criteria as well as the nature of the soliton pulses undergo significant changes as we discuss below.

We note that for finite $S$, integrating (\ref{fn_diff2}) once gives us,
\begin{equation}
\frac{1}{2} \left( \frac{\partial \phi}{\partial \zeta} \right)^2  = e^{\phi} + v_d \sqrt{v_d^2 - 2\phi} + \int_{-\infty}^{\infty}
S(\zeta)\frac{\partial \phi}{\partial \zeta} d \zeta + C
\label{sag_eq_sou}
\end{equation}
which is not amenable to a Sagdeev potential like analysis. We have therefore solved Eq.~(\ref{sag_eq}) numerically for 
a Gaussian shaped source object ($S(\zeta) = A_m \exp\left[-(\zeta/\delta)^2\right]$) and delineated the existence regions and shapes of the solitonic solutions in the parametric space of $v_d$ and the source amplitude $A_m$ and the source width $\delta$. Plots of a few such solitonic solutions with $A_m=0.1$ and $\delta=12.0$ are shown in Fig.~\ref{figure_1} for different values of $v_d$. In contrast to the pure KdV soliton or its large amplitude generalization (represented by Eq.~(\ref{sag_eq})) these pulse like solutions exist for $v_d <1$. The amplitude of these solutions increases with the increase in $v_d$ as well as with the increase in the source
amplitude $A_{m}$. This scaling is seen in the series of curves shown in Fig.~\ref{phi_max_vs_vd} in the region $v_d <1$. 

\begin{figure*}[ht]
\centering
\includegraphics[height=8.0cm,width=10.0cm]{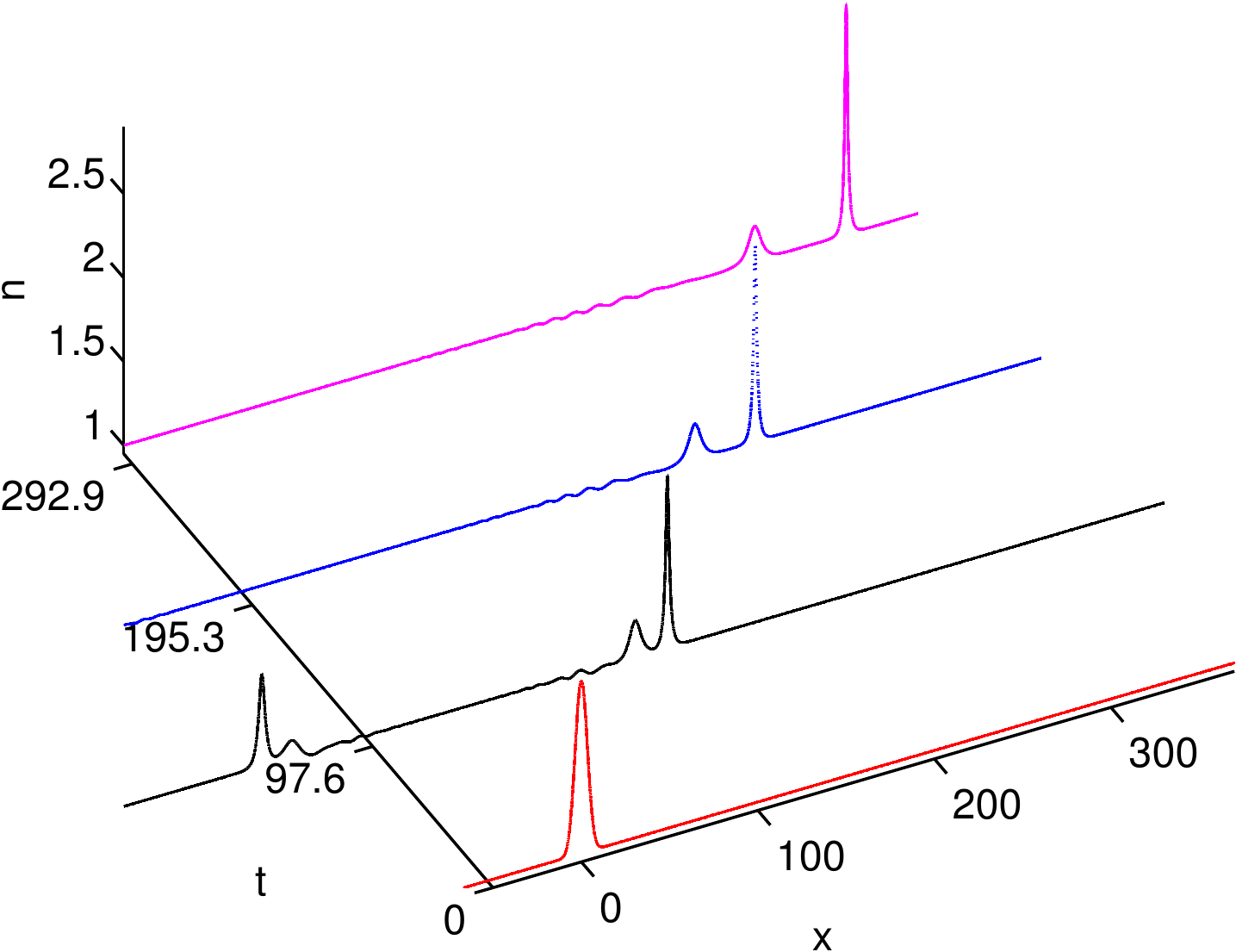}
\caption{Formation and evolution of ion acoustic solitons from an initial perturbation and in the absence of a charged moving source in the plasma. The initial profile of the stationary perturbation is of a Gaussian form. As is clearly seen, the pulse breaks up into two oppositely traveling pulses that evolve into solitons and
residual ``radiation'' structures. In the figure the right going pulse is shown for larger times.}
\label{presol_nosource}
\end{figure*}
\begin{figure}[ht]
\centering
\includegraphics[height=7.0cm,width=8.0cm]{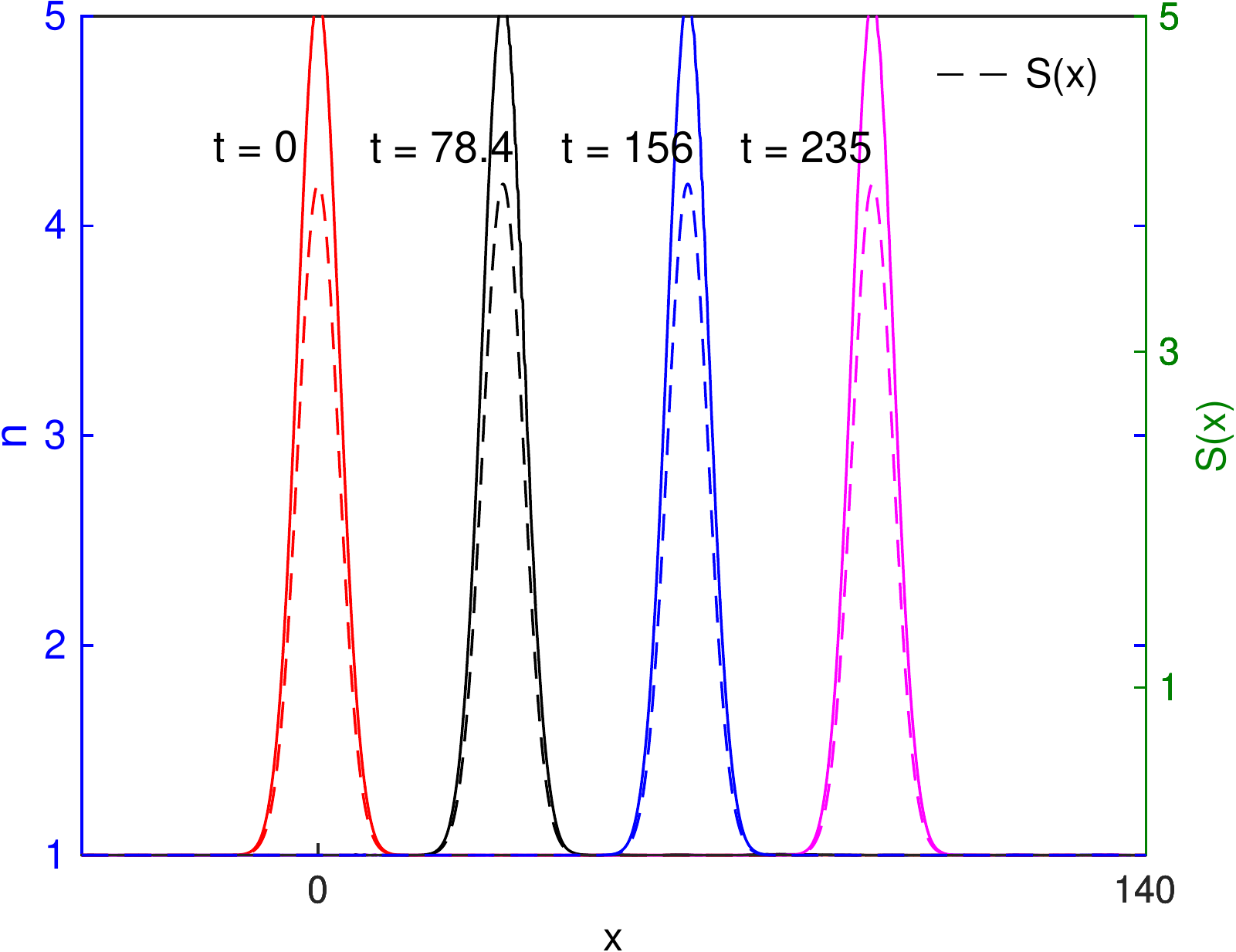}
\caption{Evolution of an exact localized solution obtained by solving Eqs.~(\ref{basiceq1}-\ref{basiceq3}) in the stationary frame with a source term of the form 
$S(\zeta)=A_m \exp\left[-(\zeta/\delta)^2\right]$ with associated parameters 
$A_m = 4.01,\delta = 4.88$ and $v_d = 0.4$. The dashed line depicts the source object density while the solid line shows the
evolution of the ion density.}
\label{exact_vd0p4_sol_evol}
\end{figure}
As the value of $v_d$ crosses unity there is a qualitative change in the shape of the stationary pulse solutions. Instead of the bell shaped structures observed for $v_d < 1 $ solutions, they begin to develop modulations near the peak. Typical solitary structures of this kind are shown in Fig.~\ref{fig_vd_gt1} for three different combinations of $v_d,\delta$ and $A_m$ representing a single peak soliton ($v_d=0.85,\delta=12,A_m=0.01$), a two peak soliton ($v_d=1.2,\delta=12,A_m=0.01$) and a three peak soliton ($v_d=1.5\delta=5.04,A_m=0.01$) respectively. 
These solutions develop increased number of modulations as any one of the parameters $v_d$, $A_m$ or $\delta$ are increased and there is also a corresponding rise in their
maximum amplitudes. This scaling behavior is shown in the curves depicted for $v_d>1$ in Fig.~\ref{phi_max_vs_vd}.
These solutions are a more generalized version of the so called ``pinned'' solitons of the fKdV equation and are valid beyond the weak nonlinearity assumption of the fKdV model. 

\begin{figure*}[ht]
\centering
\includegraphics[height=8.0cm,width=10.0cm]{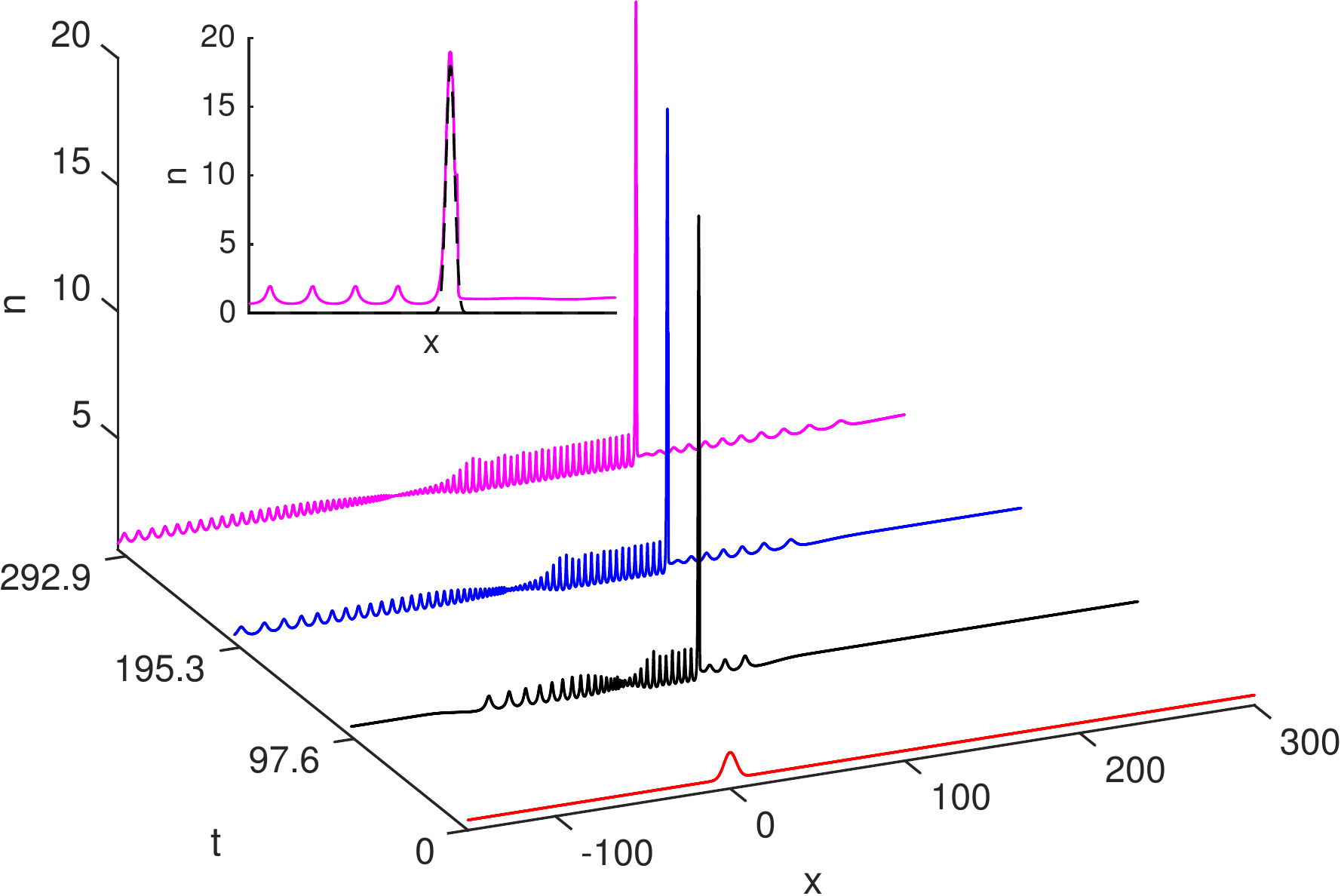}
\caption{Wake structures formed by a moving source traveling at $v_d=0.5$. Inset shows evolution of initial perturbation along with the evolution of the source}
\label{wakes}
\end{figure*}

\begin{figure*}[ht]
\centering
\includegraphics[height=8.0cm,width=10.0cm]{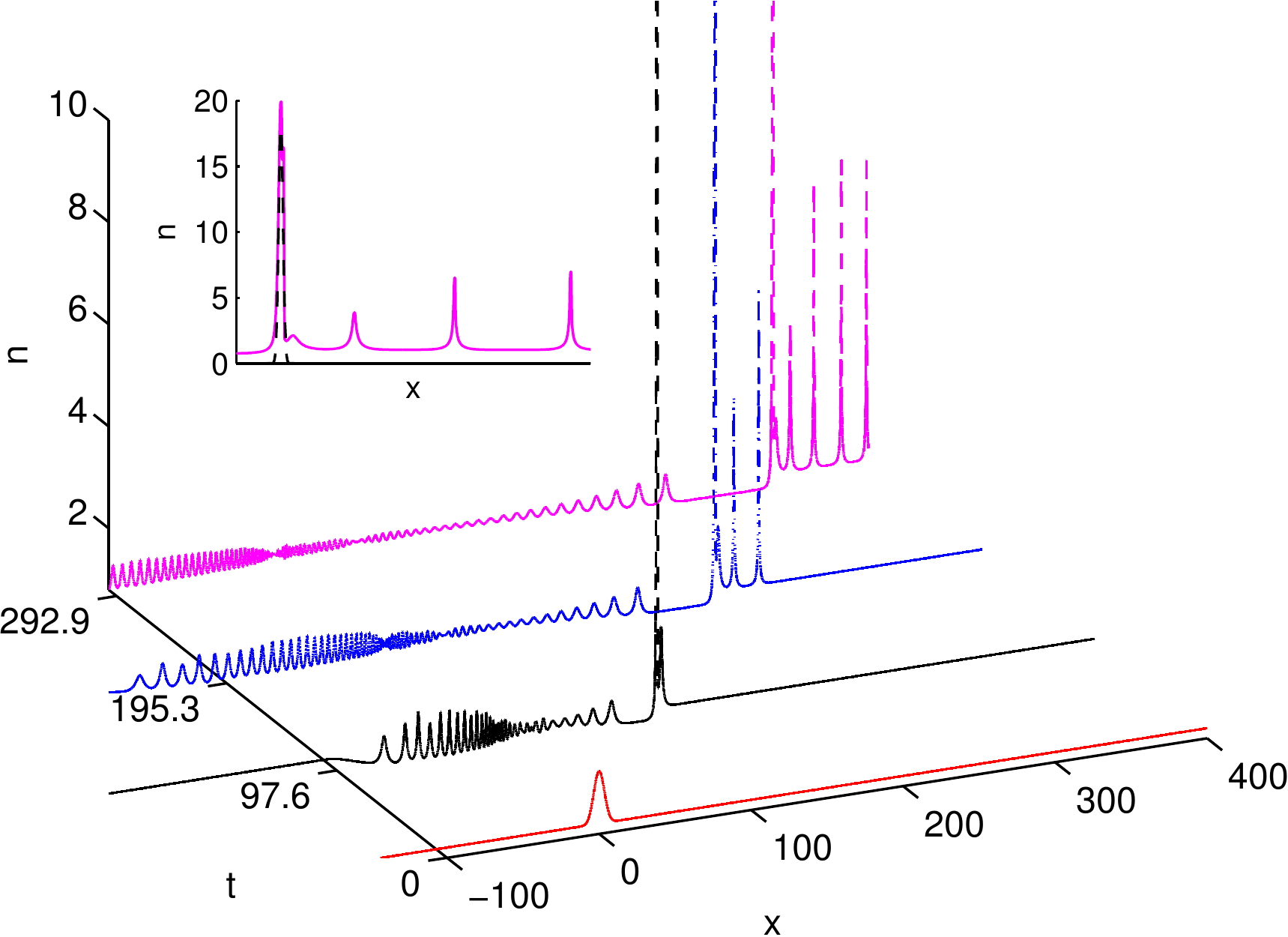}
\caption{Formation and evolution of precursor solitons due to a charged source moving at $v_d = 1.15$. The density profile of source has been
taken as $ S(\zeta) = A_m \exp\left[-(\zeta/\delta)^2\right]$ with $A_m = 18$ and $\delta = 0.488$. Inset shows evolution of initial
perturbation along with the evolving source and shows the solitons moving ahead of the source in the medium.}
\label{presol}
\end{figure*}
\section{Numerical solutions of the full set of model equations}
We now discuss the numerical solutions of the full set of model equations, namely Eqs.~(\ref{basiceq1}-\ref{basiceq3}),
in an attempt to seek generalization of the solutions known for the fKdV equation. We have used the flux corrected transport method \cite{lcpfct} to evolve the 
 continuity Eq.~(\ref{basiceq1}) and the momentum
Eq.~(\ref{basiceq2})  while Poisson's equation, Eq.~(\ref{basiceq3}) has been solved by
employing the following successive relaxation method in Fourier space, 
\begin{eqnarray}
&& \frac{\partial \phi^{n+1}}{\partial h}  = \frac{\partial^2 \phi^{n+1}}{\partial x^2} - e^{\phi^n} + n - S(x - v_d t^n) \nonumber \\
&-&\left(k_x^2 + \frac{1}{dh}\right) \phi^{n+1} = \it{ft} \left[ e^{\phi^n} - n +S(x - v_d t^n) \right] \nonumber \\
&& \phi^{n+1} = \it{ift} \left[ \left(\frac{-1}{k_x^2 + dh^{-1}}\right) \left(\it{ft} \left[ e^{\phi^n} - n +S(x - v_d t^n) \right] \right) \right]
\nonumber \\
 \label{poiss_evol}
\end{eqnarray}
Here $n$ is the index denoting time steps and $dh$ is a pseudostep used to relax the value of $\phi$ during successive iterations.
Eq.~(\ref{poiss_evol}) has been solved iteratively to get a solution for $\phi$ at each time step of the evolution of the
continuity and momentum equations. To test the efficacy and physical validity of the code we have first considered the evolution of an arbitrary perturbation in the absence of the source term i.e. in the limit of the generalized 
regime of nonlinear ion acoustic waves. The results are displayed in Fig.~\ref{presol_nosource} where it is seen that the initial stationary pulse splits into two pulses traveling in opposite directions (to conserve velocity) and eventually these pulses evolve into soliton structures. The right going and left going soliton solutions correspond to those that can be obtained by a Sagdeev potential formulation of eqns. (\ref{basiceq1}-\ref{basiceq3}) by transforming to the frames $\xi = (x - u_{ph} t)$ and $\xi = (x + u_{ph} t)$ respectively. 

\subsection{Time evolution of the generalized ``pinned'' solitons}

We next discuss the temporal behaviour of the generalized ``pinned'' soliton structures discussed in the previous section. To test the integrity and stability of these solutions we have used them as initial conditions in the full set of fluid Eqs.~ (\ref{basiceq1}-\ref{basiceq3}) and evolved them in time. In Fig.~\ref{exact_vd0p4_sol_evol}, we show the time evolution of the solitary pulse solution for the Gaussian form of source 
term $S(\zeta)=A_m \exp\left[-(\zeta/\delta)^2\right]$ with parameters 
$A_m = 4.01,\delta = 4.88$ and $v_d = 0.4$. As can be seen, the solutions evolve without distortion and loss of amplitude for several ion frequency periods suggesting that these exact time stationary numerical solutions are stable and robust structures of the full driven electron-ion plasma system. 

\subsection{Excitation of wakes and pre-cursor solitons}
\label{rand_exact}
We have next evolved different localized initial profiles which are not exact solutions of the stationary equations discussed in section \ref{ext}. To begin with, we look at the response of the system to a Gaussian source moving at a subsonic speed of $v_d=0.5$ with $A_m=18$ and $\delta=0.488$. As shown in Fig.~\ref{wakes} we see the development of prominent wake structures in the downstream region of the source. For a supersonic moving source, on the other hand, with $v_d=1.15$ and the same amplitude and width as before, the system response is dramatically different and is shown in Fig.~\ref{presol}. We notice that apart from wakes in the downstream region there are solitonic excitations in the upstream region that are moving faster than the source. These precursors exist for a range of velocities $v_d$ and cease to exist if $v_d \gtrsim 1.5$. This behaviour is analogous to what has been observed for the fKdV but in this case it emerges from the full set of fluid equations and is therefore valid for arbitrary amplitude perturbations. The results depicted in Fig.~\ref{presol} constitute one of the principal findings of our numerical investigation and provide a strong basis for expecting similar excitations to occur in a plasma system.

\section{Summary and Discussion}
\label{concl}
In this paper we have studied the evolution of wave structures excited by a moving charged body in a plasma medium. While such traveling sources are known to excite linear wake structures behind them we have investigated the phenomenon of nonlinear excitations taking place ahead of the moving body. Such excitations in the form of precursor solitons have been predicted and observed for water waves and successfully modeled by the forced KdV equation. The fKdV equation has also been shown to hold for weakly nonlinear and dispersive ion acoustic waves that are driven by a source moving in a plasma \cite{Sen2015} giving rise to the interesting possibility of their experimental realization. Our principal motivation in this work has been to try and go beyond the weakly nonlinear and dispersive limit and examine the question of the existence of precursor solitons beyond the fKdV model. For this we have considered the full set of cold fluid equations and obtained its solutions in the presence of a moving source term. Our numerical analyses reveal a rich variety of solutions including wakes, precursor solitons and the so called ``pinned'' solitons (that move with the source) lending credence to the possibility of observing such excitations in practical situations. As discussed before, such situations can arise in a variety of circumstances such as during the interaction of the ionospheric plasma with charged space objects like satellites and satellite debris, the interaction of high velocity charged beams with the ablated plasma of a target in inertial fusion schemes. One can also carry out controlled laboratory experiments using flowing plasmas passing over stationary electrostatic potentials and look for nonlinear excitations in the upstream region. Our results can prove useful in guiding the search for observing such excitations in some of the natural situations mentioned above as well as in laboratory investigation of this novel phenomenon. \\

%
\end{document}